\documentclass[a4paper, showpacs, prb, preprint, aps, superscriptaddress, amsmath, amssymb, amsfonts, floatfix]{revtex4-1} 
\usepackage[version=3]{mhchem}
\usepackage{comment}
\usepackage{graphicx}
\usepackage{latexsym}
\usepackage[caption=false]{subfig}

\begin{document}

\title{Nonlinear effects in the propagation of optically generated magnetostatic volume mode spin waves}

\author{L.J.A.\ van Tilburg}
\email{ltilburg@science.ru.nl}
\affiliation{Radboud University, Institute for Molecules and Materials,
                Heyendaalseweg 135, 6525AJ Nijmegen, The Netherlands}

\author{F.J.\ Buijnsters}
\affiliation{Radboud University, Institute for Molecules and Materials,
                Heyendaalseweg 135, 6525AJ Nijmegen, The Netherlands}

\author{A.\ Fasolino}
\affiliation{Radboud University, Institute for Molecules and Materials,
                Heyendaalseweg 135, 6525AJ Nijmegen, The Netherlands}

\author{T. Rasing}
\affiliation{Radboud University, Institute for Molecules and Materials,
                Heyendaalseweg 135, 6525AJ Nijmegen, The Netherlands}

\author{M.I.\ Katsnelson}
\affiliation{Radboud University, Institute for Molecules and Materials,
                Heyendaalseweg 135, 6525AJ Nijmegen, The Netherlands}
\date{\today}

\begin{abstract}
\textsc{As published on 24 August 2017 - prb 96, 054437 (2017)}\\
     Recent experimental work has demonstrated optical control of spin wave emission by tuning the shape of the optical pulse (Satoh et al.\ Nature Photonics, 6, 662 (2012)). We reproduce these results and extend the scope of the control by investigating nonlinear effects for large amplitude excitations. We observe an accumulation of spin wave power at the center of the initial excitation combined with short-wavelength spin waves. These kind of nonlinear effects have not been observed in earlier work on nonlinearities of spin waves. Our observations pave the way for the manipulation of magnetic structures at a smaller scale than the beam focus, for instance in devices with all-optical control of magnetism.\\
\end{abstract}

\maketitle 
\section{Introduction \label{intro}}
The study of propagating spin waves, collective excitations of the magnetization, is a subject of great interest in the emerging field of spintronics as they can be used in magnetic switching \citep{kammerer2011magnetic}, manipulation of domain walls \citep{PhysRevLett.112.147204} and logic devices \citep{hertel2004domain}. In magneto-optical pump-probe experiments, for instance in \citet{satoh2012directional}, the spin waves can be studied with both high spatial and temporal resolution. Furthermore, recent advancements in data gathering techniques\cite{hashimoto2014ultrafast} have greatly reduced measuring times to probe the spatial distribution of spin wave flow, making an experimental study of spin wave flow over a large area feasible.

One way of generating spin waves in a magnetic system is by means of the inverse Faraday effect (IFE). \citep{PhysRev.143.574, kimel2005ultrafast,PhysRevB.84.214421} In insulators with sizable spin-orbit coupling, a circularly polarized light pulse induces an effective magnetic field pulse $\mathbf{H}_{\mathrm{ind}} \propto \mathbf{E} \times \mathbf{E}^*$ where $\mathbf{E}$ is the electrical component of the incident light. The induced field is directed along the wave-vector $\mathbf{k}$ of the light. The IFE is based on an impulsive Raman scattering process and does not rely on absorption.\citep{kirilyuk2010ultrafast} Thus, it can be used to excite spin waves at high fluence without depositing any heat in the system.

Recent experiments by \citet{satoh2012directional} on the ferrimagnetic garnet \ce{Gd4/3Yb2/3BiFe5O12} have demonstrated directional control of the spin wave emission pattern. Garnets are particularly suitable for optical spin-wave experiments because many of them have a large magneto-optical response and relatively low spin wave damping.\citep{hansteen2006nonthermal} Recently, even all-optical switching of the magnetization was demonstrated in such garnet.\cite{stupakiewicz2017ultrafast} Since the magnetization couples so strongly to the optical field, garnets are ideal to study the nonlinear effects of spin wave propagation, such as self-focusing: the concentration of the spin wave power on a single point. 

Nonlinear effects of magnetostatic spin waves have a relatively low excitation threshold due to the low value of the ratio of dispersion and diffraction. \citep{PhysRevLett.81.3769,Demokritov2001441} Previous work on nonlinear spin wave dynamics found self-channeling of the spin waves for a flow along a (quasi)-one-dimensional sample.\citep{bauer1997direct,boyle1996nonlinear,demidov2009nonlinear} For spin waves propagation in two dimensions, Brillouin light-scattering experiments showed spin wave focusing into quasi-stable \textquoteleft bullets\textquoteright. \cite{PhysRevLett.81.3769,buttner2000linear, Demokritov2001441}  In these experiments, the spin waves were excited by a microwave antenna, creating a spectrally narrow wave packet with a well defined propagation direction.
\newpage
Here, we investigate the propagation of \emph{optically} generated spin waves. Due to the nature of this excitation, many wavenumbers are excited and the spin waves will propagate in all in-plane directions. Such a spin wave distribution has not been studied before in the nonlinear regime. We investigate the dynamics of an optically excited wave packet by numerically solving the Landau-Lifshitz-Gilbert equation for a suitable initial configuration. This approach makes a connection with recent magneto-optical experimental work\cite{satoh2012directional,shen2015laser,au2013direct} and allows us to extend the current understanding to the nonlinear regime.

This article is organized as follows. In Sec.\ \ref{method} we introduce the system under study and explain our numerical approach. In Sec.\ \ref{results} we present numerical results of the spin wave propagation. Sec.\  \ref{part_lin} deals with the linear regime (small initial excitations) where one can derive the spin wave dispersion (\ref{part_disp}). We also present typical spin wave patterns that an optical pulse excites (\ref{part_lin_patterns}) and compare them to experimental data.  Sec.\ \ref{part_nonlin} expands the scope to large initial excitations and presents results for nonlinear spin wave dynamics. In Sec.\ \ref{discuss} we summarize our results and discuss the experimental feasibility of these nonlinear effects.

\section{Method \label{method} }
We study the spin wave pattern that emerges when we prepare our system in an initial configuration at $t=0$ and calculate the time evolution by integrating the Landau-Lifshitz-Gilbert equation (LLG)\citep{Landau1935},
\begin{align}
   \frac{\partial \mathbf{m}}{\partial t} = - \gamma \left( \mathbf{m} \times \mathbf{H}_{\mathrm{eff}} \right) + \frac{\alpha}{m} \left( \mathbf{m} \times  \frac{\partial \mathbf{m}}{\partial t} \right) \ , \label{llg}
\end{align} 
where $\mathbf{m}$ is defined as the magnetization vector per volume, $\gamma$ is the gyromagnetic ratio, $\alpha$ is the Gilbert damping parameter and $\mathbf{H}_{\mathrm{eff}}$ is the effective magnetic field, calculated as the functional derivative of the energy with respect to the magnetization. The first term on the r.h.s.\ of Eq.\ \ref{llg} describes a precessional motion of the magnetization around the effective magnetic field. The second term describes the damping of this precession, ultimately aligning the magnetization to the field.
\newpage
We consider the following Hamiltonian 
\begin{align}
   \mathcal{H}_\mathrm{tot} = \mathcal{H}_\mathrm{ani} + \mathcal{H}_\mathrm{dip} + \mathcal{H}_\mathrm{Z}~, \label{hamiltonian}
\end{align}
for a uniform slab of thickness $L$ of a material with out-of-plane easy-axis magnetic anisotropy \makebox{$\mathcal{H}_\mathrm{ani} \sim K (\mathbf{m} \cdot \hat{\mathbf{e}}_\mathrm{z})^2$} where $\hat{\mathbf{e}}_\mathrm{z}$ is the normal to the surface. Magnetostatic dipole-dipole interactions are taken into account in the $\mathcal{H}_\mathrm{dip}$ term and $\mathcal{H}_\mathrm{Z} \sim M_S H_\mathrm{Z}^x$ represents the Zeeman energy for an in-plane static external magnetic field that aligns the equilibrium magnetization $M_S$ along $\hat{\mathbf{e}}_\mathrm{x}$. For garnets, the exchange length, $l_\mathrm{ex} = \sqrt{2A/\mu_0 M_S^2}$ with $A$ the exchange constant, is of the order of nanometers, very small compared to the typical wavelengths of magnetostatic waves. \citep{edwards2013magnetostatic}  Therefore, we neglect the contribution of exchange. This defines the applicability of our approach as long as the wavelength $2\pi/k$ of the spin waves is large compared to $l_\mathrm{ex}$ i.e.\ $kl_\mathrm{ex} \ll 1$.

We solve Eq.\ \ref{llg} on a 2400 $\times$ 2000 square lattice of macrospins with lattice parameter $a=1\mu$m  and periodic boundary conditions.\footnote{We are interested in the spin wave propagation in a 2 ns interval. The size of the lattice ensures that the spin waves do not reach the boundaries of the lattice during this interval} We fix the spin wave profile in the direction normal to the plane to be uniformly distributed (the uniform mode approximation). This approximation enables us to simulate the system as effectively two dimensional, greatly decreasing computational costs. Since the lowest order profile of the volume modes is homogeneous throughout the sample, this approximation is also effective beyond the thin-film limit\cite{frankB_disp}. The thickness $L$ of the sample defines the intrinsic length scale in the calculation. To be able to capture the magnetization dynamics, we have taken care that the resolution is much smaller than this length scale ($a \ll L$). Note that an optical excitation relying on the IFE acts uniformly through the sample, since it can be tuned to operate at an optical wavelength where the system is transparent.

\begin{figure}[hbt]
\centering
\hfill\null
\subfloat[\label{perspective view}]{
        \includegraphics[width=.5\textwidth]{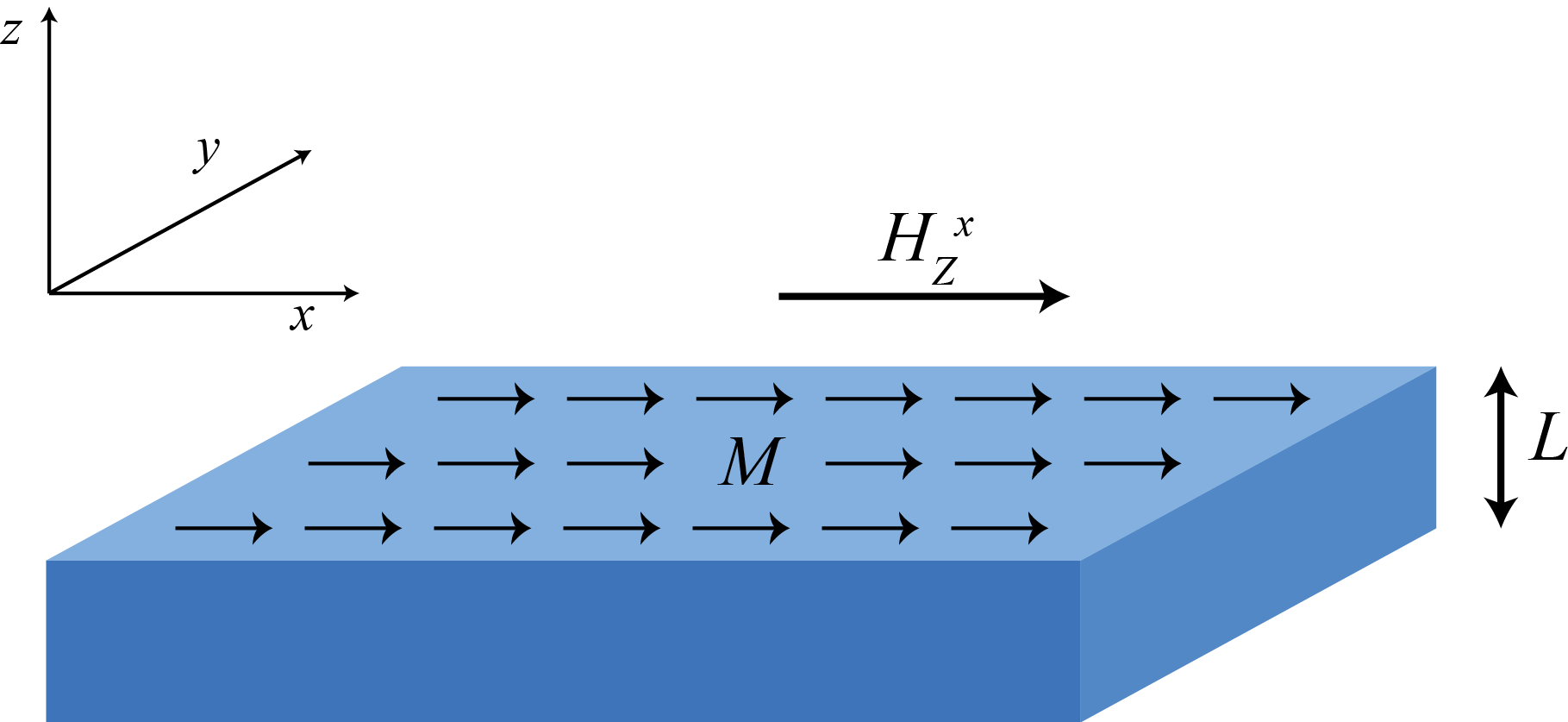}  
} \hfill
\subfloat[\label{top_view}]{
        \includegraphics[width=.32\textwidth]{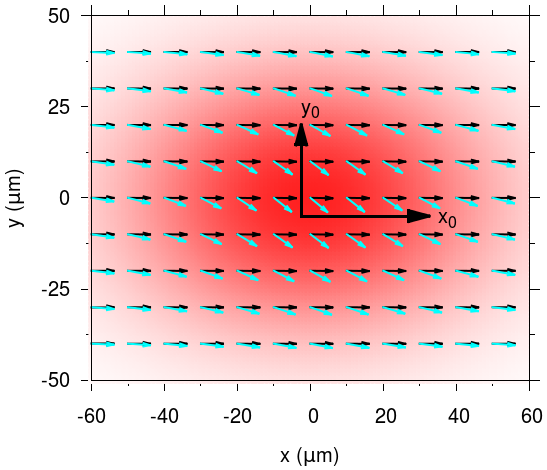}
}
\hfill\null
\caption{A slab of material of thickness $L$ with an external field $H^x_Z$ along the $x$ axis to align the in-plane magnetization: \protect\subref{perspective view} system geometry and state before illumination; \protect\subref{top_view} top view of the excited state of the in-plane magnetization after a pulse with $x_0=35\ \mu\mathrm{m}$ and $y_0=25\ \mu\mathrm{m}$ and varying laser intensities, according to Eq.\ \ref{initial_eq}. The red background shows the intensity profile of the optical field. The small vectors show the magnetization for $\vartheta=1^\circ$ (black) and $\vartheta=40^\circ$ (light blue). \label{initial_conf}}
\end{figure}

Since the period of the typical spin waves in our system is much longer than the pulse duration of the incident light, the effect of the optical field can be considered as instantaneous. Thus, in accordance with Eq.\ \ref{llg}, for light at normal incidence the IFE leads to a tilt of the magnetization over an angle $\theta$ in the direction of $-(\mathbf{m} \times -\hat{\mathbf{e}}_\mathrm{z})$ i.e.\ a clockwise rotation around $\hat{\mathbf{e}}_\mathrm{z}$. The profile of the optical pulse determines the spatial extent of the tilted spins. 
\newpage
In this paper we consider the tilt angle profiles having the Gaussian shape
\begin{align}
   \theta(x,y) = - \vartheta \exp \left( -\frac{x^2}{2x_0^2} - \frac{y^2}{2y_0^2}  \right) \ ,
\end{align}
where $x_0$ and $y_0$ determine the ellipticity of the laser spot, $\vartheta$ is the maximum amplitude and the minus sign gives the rotation direction.  Pulses with $x_0 \neq y_0$ can be achieved by using an aperture as is done in Ref.\ \onlinecite{satoh2012directional}.  A strong induced field due to the IFE effect would result in a large value for $\vartheta$.
Since $|\mathbf{m|}/M_S=1$, the local magnetization after the pulse is given by
\begin{align}
   \mathbf{m}_{\vartheta}(x,y) &= \cos\theta~ \hat{\mathbf{e}}_\mathrm{x}  + \sin \theta ~\hat{\mathbf{e}}_\mathrm{y} \ .  \label{initial_eq}
\end{align}
  Figure \ref{initial_conf} shows the systems geometry and two excited states resulting from two laser pulses for different values of $\vartheta$.

The calculations were performed with the micromagnetics code developed in our group.\cite{PhysRevLett.116.147204} The numerical integration follows the implicit midpoint scheme (IMP) which conserves the magnitude of the initial spins.\cite{0953-8984-22-17-176001} A second property of this integration scheme is that it exactly conserves energy for Hamiltonians that are of quadratic order in the magnetization when damping is zero. \cite{d2006midpoint} 

\clearpage
\section{Results \label{results} }
\subsection{The linear regime $\left( \vartheta=1^\circ \right)$  \label{part_lin}}
\subsubsection{The uniform mode dispersion \label{part_disp}}
If we assume a uniform spin wave profile throughout the depth of the material, the dipole-dipole term in the Hamiltonian of Eq.\ \ref{hamiltonian} can be treated analytically. \cite{hurben1996theory,kalinikos1980excitation,frankB_disp} When the amplitude of the initial excitation $\vartheta$ is small, it is possible to linearize the Landau-Lifshitz equation without damping and calculate the dispersion $\omega$ of a spin wave with wave-vector $\mathbf{k}$
\begin{align}
   \omega (\mathbf{k}) &= \mu_0 |\gamma| \sqrt{ \left( H^x_\mathrm{Z} + M_x \frac{k_y^2}{k^2} \frac{e^{-kL}+kL-1}{kL} \right) \left( H^x_\mathrm{Z} + M_x \frac{1-e^{-kL}}{kL} -\frac{2K}{\mu_0 M_S} \right)} \ , \label{disp_eq}
\end{align}
with $k^2= k_x^2 + k_y^2$ and $M_x$ is the component of the magnetization along the x-direction (since $\vartheta$ is small, $M_x\approx M_S$). For spin waves parallel to the equilibrium magnetization ($\mathbf{k} \parallel \mathbf{M}$) the dispersion reduces to the well known result for volume spin waves in thin films. \citep{Demokritov2001441,stancil2009spin}

\begin{figure}[hbt]
\centering
\subfloat[\label{disp_map}]{
        \includegraphics[width=.4\textwidth]{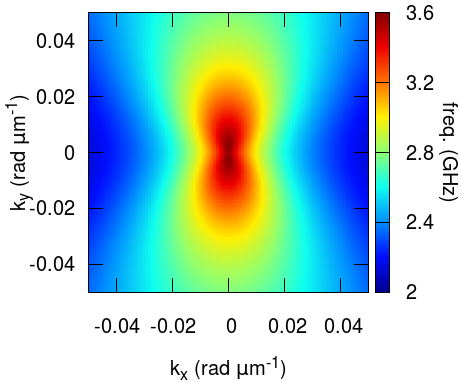}  
}
\subfloat[\label{disp_graph}]{
        \includegraphics[width=.5\textwidth]{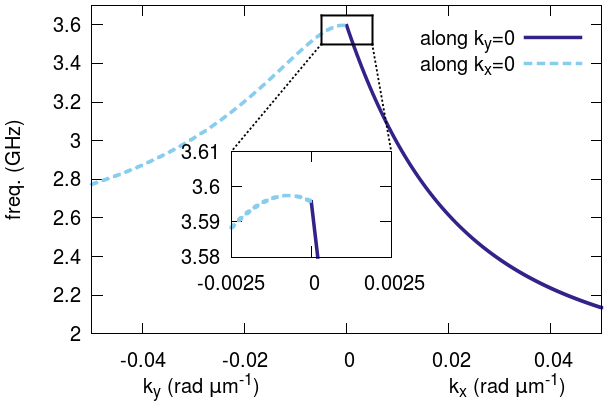}
}
\caption{The dispersion relation for magnetostatic volume mode spin waves in the xy-plane according to Eq. \ref{disp_eq} using parameters from Table \ref{parameters}: \protect\subref{disp_map} shows the dispersion in the full 2D plane; \protect\subref{disp_graph} shows a cut through k-space along $k_x$ (solid dark blue line) and $k_y$ (dashed light blue line). The waves propagating along the external field $H^x_\mathrm{Z}$ all have a negative group velocity. Waves propagating perpendicular to the field have a different behavior for small $k$.  \label{disp_all}}
\end{figure}

Figure \ref{disp_all} shows the dispersion relation of Eq.\ \ref{disp_eq} along the x- and y-axis, using the parameters from Table \ref{parameters}. 

\newpage
For almost all wave-vectors $|\mathbf{k}|$, the dispersion has negative slope. The associated spin waves are called backward volume magnetostatic waves (BVMSW) due to their negative group velocity. At $k=0$, the dispersion has as discontinuity in its derivative, indicative of long-range interactions.  For completeness we note that for very  small values of $k_y$,  the dispersion starts with a positive slope (see inset Fig.\ \ref{disp_graph}), changing sign when $k_y$ increases. This maximum in the dispersion results from the uniform mode approximation and disappears when the spin wave profile is allowed to vary in the z-direction.\cite{frankB_disp} In that case, the dispersion starts out flat for small $k_y$, decreasing when $k_y$ increases.

\subsubsection{Spin wave patterns \label{part_lin_patterns} }
\begin{figure}[hbt]
\centering
\subfloat[\label{lin_circ}]{
        \includegraphics[width=.3\textwidth]{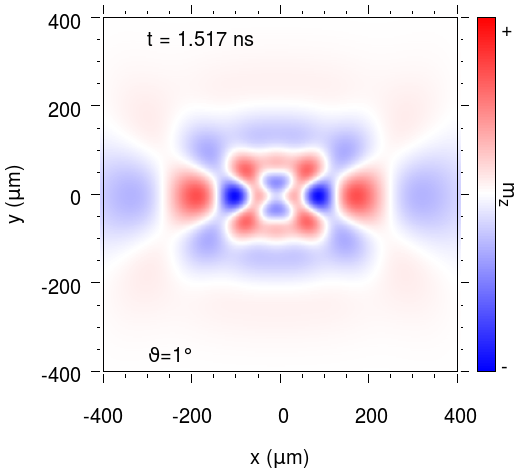}
} \quad
\subfloat[\label{lin_ellipY}]{
        \includegraphics[width=.3\textwidth]{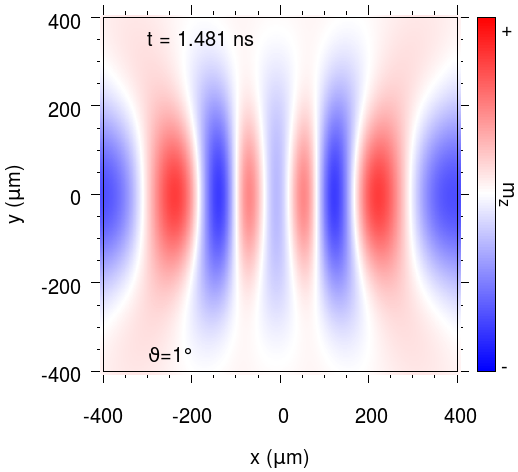}  
} \quad
\subfloat[\label{lin_ellipX}]{
        \includegraphics[width=.3\textwidth]{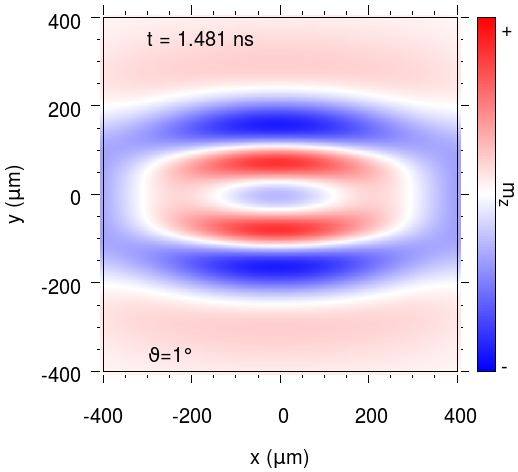}
} \quad
\caption{Three snapshots of the spin wave pattern calculated at $t = 1.5$ ns: \protect\subref{lin_circ} with a circular initial excitation $x_0=y_0=25 \mu$m (top left); \protect\subref{lin_ellipY} with an elliptical initial excitation $x_0= 35 \mu$m, $y_0=140 \mu$m; \protect\subref{lin_ellipX} with $x_0 =140 \mu $m, $y_0=35 \mu$m. These values have been chosen so that the spin wave patterns can be compared to data in Ref.\ \onlinecite{satoh2012directional}  \label{sw_compare_to_exp} }
\end{figure}

The propagation of optically generated spin waves in the linear regime has recently been experimentally studied using a pump-probe technique by \citet{satoh2012directional}. To assert the validity of our numerical scheme, which relies on the uniform mode approximation, we have calculated the spin wave  patterns for the experimental parameters from Ref.\ \onlinecite{satoh2012directional}. 
Figure \ref{sw_compare_to_exp} shows three snapshots of the spin wave pattern at $t = 1.5$ ns for the excited state given by Eq.\ \ref{initial_eq} for $\vartheta = 1^\circ$. The size and shape of the pump spot is varied using the parameters $x_0$ and $y_0$. There is very good agreement with Figures 2a,b and 3b,c,e,f from Ref.\ \onlinecite{satoh2012directional}, confirming that our numerical scheme works well in calculating the spin wave propagation. In the Supplemental Material \cite{movies} we show a video of the dynamics of the spin wave pattern after illumination of the sample with a circular spot $x_0=y_0=25\ \mu m$ and $\vartheta=1^\circ$. Note that the calculations of spin wave patterns presented in Ref.\ \onlinecite{satoh2012directional} use a different method from the one employed by us. In Ref.\ \onlinecite{satoh2012directional} the instantaneous magnetization is calculated by integrating the Fourier transform of the initial excitation in time. In this integral the lowest order mode of the spin wave dispersion was used as input. Our scheme does not use the dispersion as an input but fixes the profile of the spin waves in the material.

\begin{table}[hbt]
\begin{tabular}{c}
\hline \hline
$\begin{aligned}
M_S &= 83\ \mathrm{kA/m} \qquad \gamma = 1.76 \cdot 10^{11}\ \mathrm{rad\ T^{-1}s^{-1}} \\
H_\mathrm{Z}^x  &= 99\ \mathrm{kA/m}\qquad \alpha = 0.02 \\
H_\mathrm{ani} &= 77\ \mathrm{kA/m} \qquad L = 110\ \mu \mathrm{m}
\end{aligned}$ \\
\hline \hline
\end{tabular}
\caption{Numerical parameters used in the simulations. The value of the anisotropy field is a factor two higher than the experimental value listed in Ref.\ \onlinecite{satoh2012directional}. The need for rescaling of the anisotropy can be attributed to the uniform mode approximation.\cite{frankB_disp} \label{parameters}}
\end{table}
\newpage

\clearpage
\subsection{The nonlinear regime $ \left( \vartheta > 1^\circ \right)$ \label{part_nonlin} }
\begin{figure}[hbt]
\centering
   \includegraphics[width=.45\textwidth]{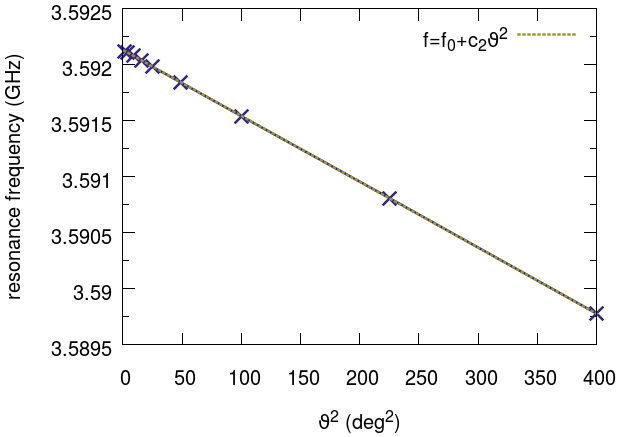}
\caption{Frequency $f=\omega/2\pi$ of the resonance mode for a lattice of $100\times100$ spins (blue crosses) plotted versus $\vartheta^2$ in the range $[0,20^\circ]$ and a linear fit to the data with slope $c_2\approx -5.87\times10^{3}$ Hz (dotted green line).  \label{fmr_vs_theta}}
\end{figure}

Since our method does not rely on a priori knowledge of the dispersion, it enables us to explore the large $\vartheta$ regime. As a first indicator of the emergence of nonlinear effects, we look at the frequency $f=\omega/2\pi$ of the resonant mode ($\mathbf{k}=0$) for increasing values of $\vartheta$. From Eq.\ \ref{disp_eq} one can see that for increasing $\vartheta$, the resonance frequency is expected to soften since the component of the magnetization along the magnetic field gets smaller for larger opening angles.\cite{khivintsev2010nonlinear} In Fig. \ref{fmr_vs_theta} we show numerical results for the resonance frequency as a function of $\vartheta^2$. The data are fitted by $f(\vartheta^2) = f_0 + c_2 \vartheta^2$, with $c_2\approx -5.87\times10^{3}$ Hz. 

\begin{figure}[htb]
\centering
\subfloat[\label{circ_theta1}]{
   \includegraphics[width=.42\textwidth]{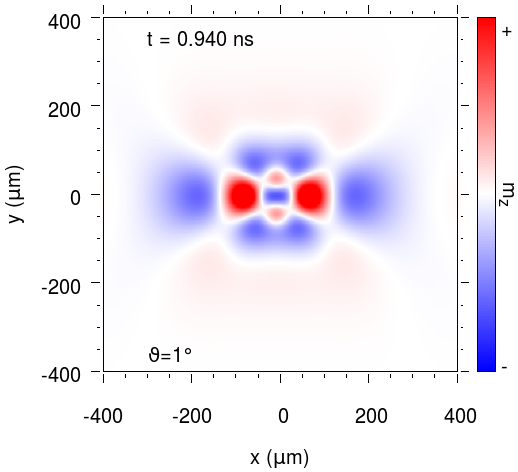} 
} \qquad
\subfloat[\label{circ_theta10}]{
   \includegraphics[width=.42\textwidth]{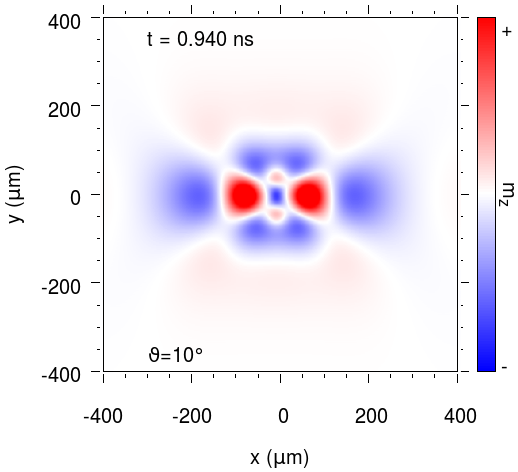} 
}\\
\subfloat[\label{circ_theta20}]{
   \includegraphics[width=.42\textwidth]{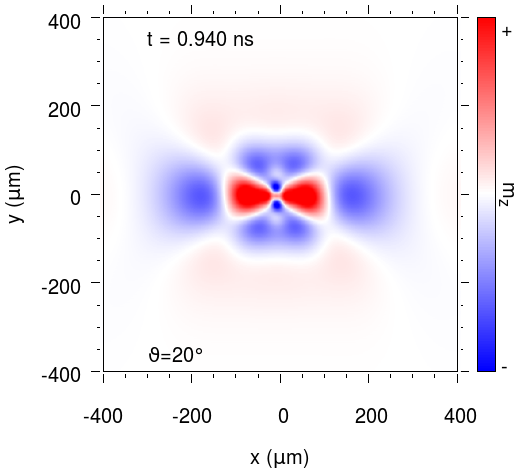} 
}  \qquad
\subfloat[\label{circ_theta25}]{
   \includegraphics[width=.42\textwidth]{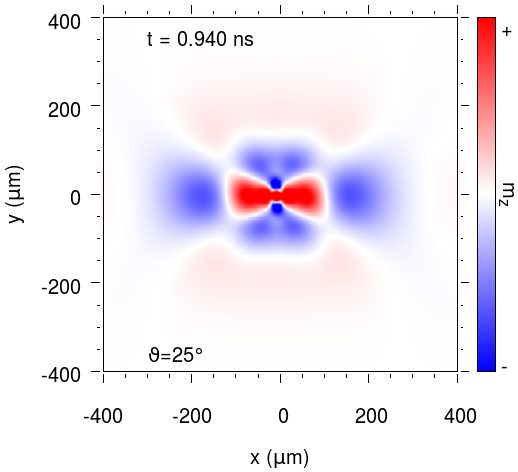}
}\\
\subfloat[\label{circ_theta30}]{
   \includegraphics[width=.42\textwidth]{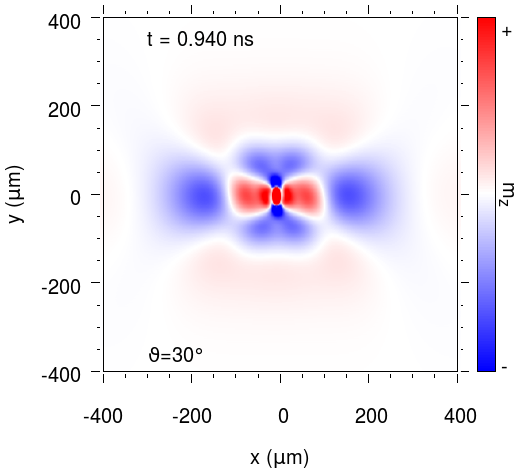}  
} \qquad
\subfloat[\label{circ_theta40}]{
   \includegraphics[width=.42\textwidth]{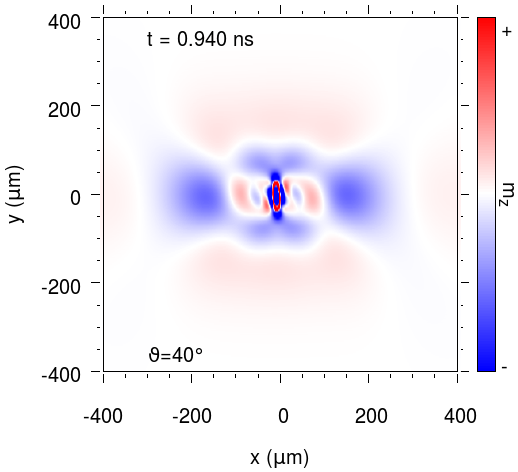}  
}
\caption{Six snapshots of the spin wave pattern observed at $t = 0.94$ ns with a circular initial excitation $x_0=y_0=|\mathbf{r}_0| = 25\mu$m and various values of $\vartheta$: \protect\subref{circ_theta1} $\vartheta=1^\circ$; \protect\subref{circ_theta10} $\vartheta=10^\circ$; \protect\subref{circ_theta20} $\vartheta=20^\circ$; \protect\subref{circ_theta25} $\vartheta=25^\circ$; \protect\subref{circ_theta30} $\vartheta=30^\circ$; \protect\subref{circ_theta40} $\vartheta=40^\circ$. The color coding is scaled to one quarter of maximum of the initial amplitude of the tilt $\sin \vartheta$ to highlight the differences in spacial profile of the spin waves. For increasing values of $\vartheta$ the observed spin wave pattern features small wavelength modes near $\mathbf{r}=0$.\label{sw_patterns_nl} }
\end{figure}

\begin{figure}[hbt]
\centering
\subfloat[\label{lineplot_theta1}]{
   \includegraphics[width=.45\textwidth]{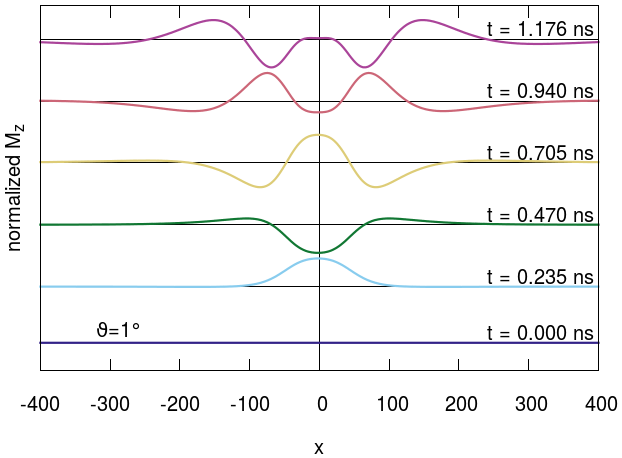} 
} \qquad
\subfloat[\label{lineplot_theta20}]{
   \includegraphics[width=.45\textwidth]{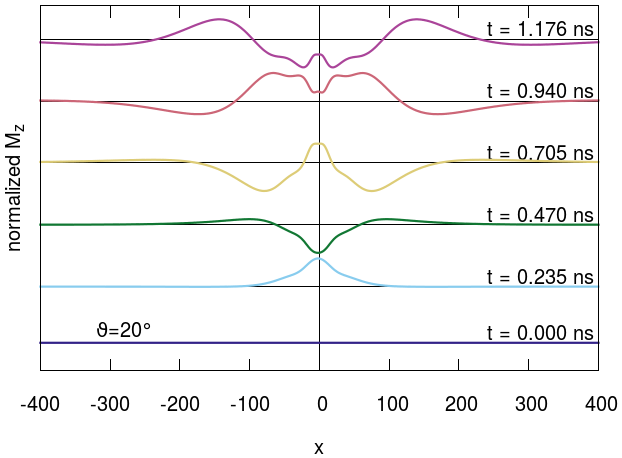}
} \qquad
\subfloat[\label{lineplot_theta25}]{
   \includegraphics[width=.45\textwidth]{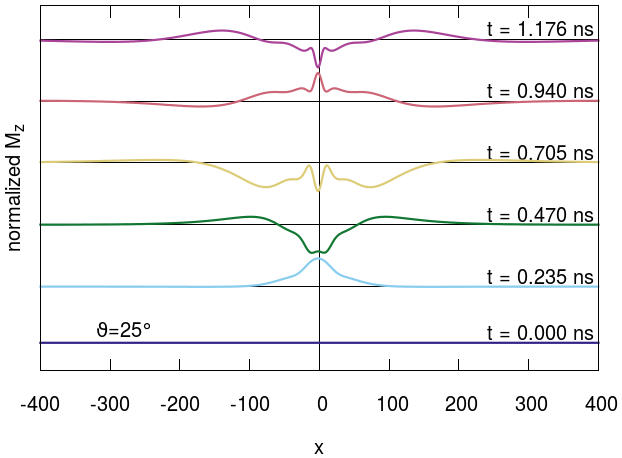}
} \qquad
\subfloat[\label{lineplot_theta40}]{
   \includegraphics[width=.45\textwidth]{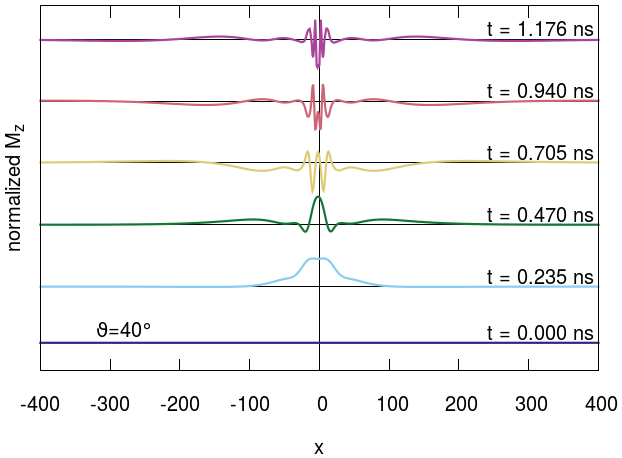}
}
\caption{The spin wave profile (z-component) along $y=0$ at different times, after an initial circular excitation with $x_0=y_0=|\mathbf{r}_0|=25\mu m$ and various values of $\vartheta$: \protect\subref{lineplot_theta1}  $\vartheta=1^\circ$; \protect\subref{lineplot_theta20}  $\vartheta=20^\circ$ \protect\subref{lineplot_theta25};  $\vartheta=25^\circ$; \protect\subref{lineplot_theta40}  $\vartheta=40^\circ$. The spin wave amplitude is normalized to aid comparison. At t=0.0 ns, the spins do not yet have a component in the z-direction. It can be seen clearly that for larger values of $\vartheta$, the spin wave power accumulates at the center of the excitation i.e.\ $x=0$. For $\vartheta=40^\circ$ this effect is quite dramatic. \label{sw_nonlin_cuts}}
\end{figure}

We now return to excitations with a spacial profile given by Eq.\ \ref{initial_eq}. In Fig.\ \ref{sw_patterns_nl} we show snapshots of the spin wave pattern at $t = 0.94$ ns for six values of $\vartheta$ and $x_0=y_0=25\ \mu m$. In the Supplemental Material \cite{movies} we show a video of the dynamics of the spin wave pattern for $\vartheta=25^\circ$. In Fig. \ref{sw_nonlin_cuts} we show the spin wave signal along the line $y=0$ for $\vartheta=1^\circ,~\vartheta=20^\circ~\vartheta=25^\circ~\mathrm{and}~\vartheta=40^\circ$ at various times. 

We observe a strong dependence on the tilt angle $\vartheta$. For low values of $\vartheta$, the spin waves propagate away from the center of the initial excitation ($x=y=0$). For larger $\vartheta$ the spin waves stay localized at $x=y=0$. Furthermore, the spin waves observed at larger $\vartheta$ have a shorter wavelength compared to the spin waves observed at the same time for lower $\vartheta$. This phenomenon becomes clearly noticeable for $\vartheta \gtrsim 20^\circ$.
\newpage
One can also notice that the symmetry of the spin wave pattern changes for larger $\vartheta$. This has to do with the asymmetry introduced due to the initial excitation given by Eq.\ \ref{initial_eq} which is negligible for small values of $\vartheta$ and becomes appreciable at larger $\vartheta$.

\begin{figure}[hbt]
\centering
\subfloat[\label{Imz_spot}]{
   \includegraphics[width=.45\textwidth]{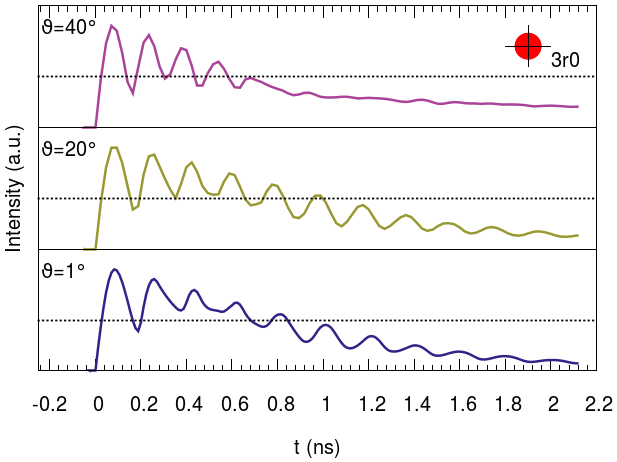}
} \qquad
\subfloat[\label{Imz_ring}]{
   \includegraphics[width=.45\textwidth]{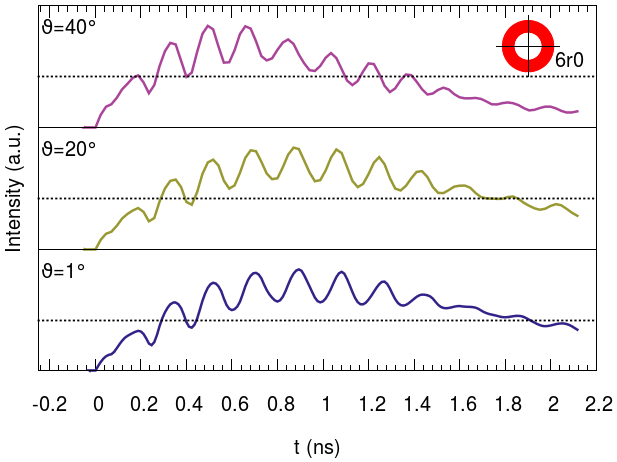}
}
\caption{Integrated spin wave power for two areas (shown in the top right corner in red) on the sample: \protect\subref{Imz_spot} a circle with radius $3r_0$ at $x=y=0$; \protect\subref{Imz_ring} a ring with $3r_0 \leq r \leq 6r_0$ centered at $x=y=0$. The peaks of the SW power have been normalized to aid comparison, the dotted lines indicate half of the maximum. The spin wave power flows outwards with increasing time. For calculations with a large initial $\vartheta$, the spin wave power in the initial circle $r \leq 3r_0$ never vanishes and actually becomes bigger than the spin wave contribution in the ring.  \label{Imz_time}}
\end{figure}

A different way to visualize the localization of spin waves is shown in Fig.\ \ref{Imz_time}. Here we plot the integrated spin wave power $\int_S |m_z| d\mathbf{s}$ for two areas $S$ on the sample, depicted in the top right corner of the figure. The spin wave power starts localized at the center of the sample within a circle with radius $r \leq 3r_0$. 

\noindent The spin waves then flow outwards as can be seen by an increase in the spin wave power in the ring $3r_0 \leq r \leq 6r_0$. For low values of $\vartheta$, the spin wave power at the initial location vanishes whereas for larger initial $\vartheta$ we see that there remains a sizeable contribution of the spin wave power within the circle $r \leq 3r_0$. For the larger $\vartheta$, the spin wave power within the circle with $r\leq 3r_0$ dominates whereas in the low $\vartheta$ case, it has propagated out of this area.

A good way to compare the spin wave flow for various $\vartheta$, is to inspect the time when the spin wave power drops below half of the maximum intensity. In the region $3r_0 \leq r \leq 6r_0$, in Fig. \ref{Imz_ring}, this happens earlier for increasing $\vartheta$. Combined with the fact that the spin wave power inside the area $ r \leq 3r_0$ in Fig. \ref{Imz_spot} stays relatively high we conclude the spin waves accumulate at the center and we may speak of self focusing.

\begin{figure}[htb]
\centering
\subfloat[\label{kmodes_theta1_ini}]{
   \includegraphics[width=.3\textwidth]{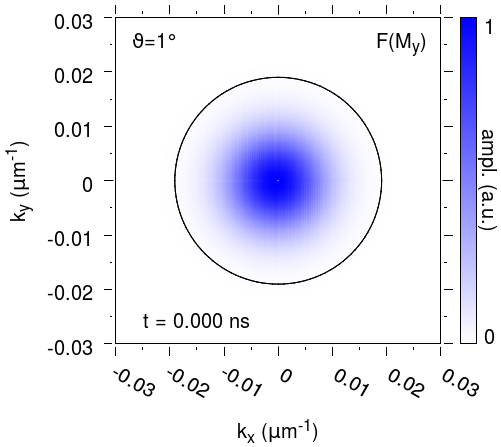} 
} \qquad
\subfloat[\label{kmodes_theta1}]{
   \includegraphics[width=.3\textwidth]{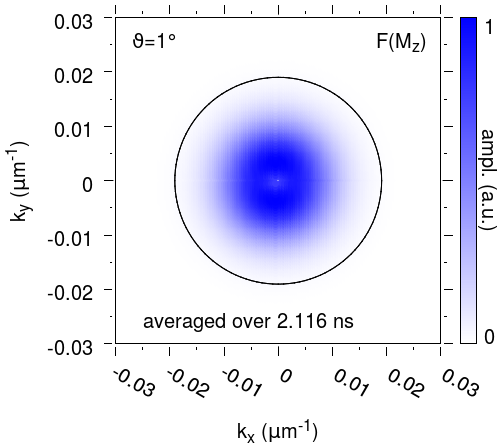} 
} \qquad
\subfloat[\label{kmodes_theta20}]{
   \includegraphics[width=.3\textwidth]{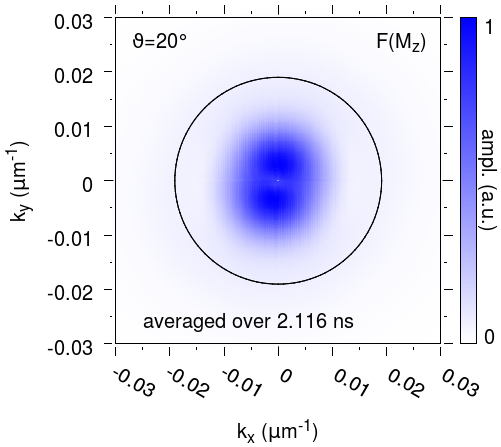}
} \qquad
\subfloat[\label{kmodes_theta40}]{
   \includegraphics[width=.3\textwidth]{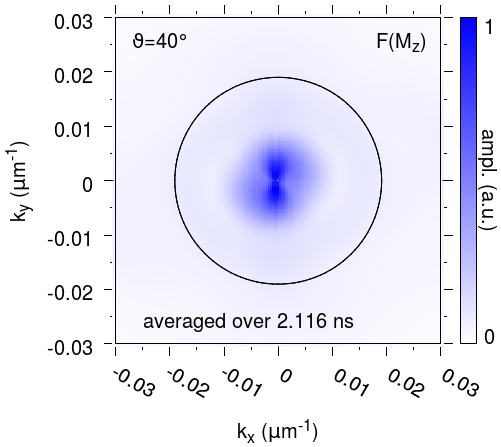}
}
\caption{The time-averaged intensity of the Fourier transform of the spin wave pattern for a circular initial excitations with $x_0=y_0=|\mathbf{r}_0|=25\mu m$ and various values of $\vartheta$: \protect\subref{kmodes_theta1_ini} the initial state for all  $\vartheta$, where the magnetization has only nonzero components in $m_y$; \protect\subref{kmodes_theta1}  $F(M_z)$ for $\vartheta=1^\circ$; \protect\subref{kmodes_theta20} $F(M_z)$ for   $\vartheta=20^\circ$; \protect\subref{kmodes_theta40} $F(M_z)$ for $\vartheta=40^\circ$. The black circle represents k=3$k_0$ and encloses the area where 99.7\% of the modes present in the initial excitation lie. For larger values of $\vartheta$ modes beyond $k=3k_0$ become populated. These modes are associated with short wavelength spin waves. \label{sw_kspace}}
\end{figure}

\newpage
To understand this effect we examine the spectral composition of the spin waves distribution. For circular excitations the Fourier transform of the initial state is a Gaussian with radius $|\mathbf{k}_0| = 1/(2\pi |\mathbf{r}_0|)$ in k-space. Each mode will precess with frequency $\omega(\mathbf{k})$ around the external field $H_\mathrm{Z}^x$. In the linear regime, the spins do not interact and the initial wavenumber distribution will not change with time. However, if $\vartheta$ increases and the system enters the nonlinear regime, modes not present in the initial excitation can become populated. This is shown in Fig.\ \ref{sw_kspace} where the time-averaged intensity of the Fourier modes associated with $M_z$ ($F(M_z)$) is plotted for three values of $\vartheta$.

\newpage
The initial wavenumbers are bounded by a circle with radius $k=3k_0$. For $\vartheta=1^\circ$, these modes do not expand outside this bound during the time evolution. However, for larger values of $\vartheta$, the spin wave modes are not confined to this initial area. Modes with higher wavenumbers arise that correspond to the short wavelength spin waves shown in Fig.\ \ref{sw_patterns_nl} and Fig.\ \ref{sw_nonlin_cuts}. The shape of the wavenumber distribution for large $\vartheta$ is reminiscent of the dispersion shown in Fig. \ref{disp_all}. This suggests that, even in the large $\vartheta$ regime where a derivation of the dispersion is not possible, the overall shape will be similar.\\

\begin{figure}[hbt]
\centering
\includegraphics[width=.6\textwidth]{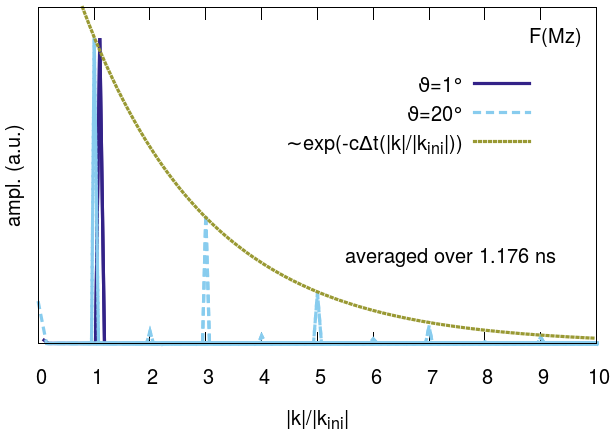}
\caption{Amplitude of $M_z(\mathbf{k})$ along $\mathbf{k}_\mathrm{ini}$ averaged over $\Delta t=1.176$ns, approximately 2.5 periods of precession, for $\vartheta=1^\circ$ (solid dark blue line) and $\vartheta=20^\circ$ (dashed light blue line). For easy comparison, we have normalized both spectra to the $\mathbf{k}_\mathrm{ini}$ peak and shifted the data for $\vartheta=1^\circ$ slightly to the right. For $\vartheta=1^\circ$, the mode with initial wave-vector $\mathbf{k}_\mathrm{ini}$ remains the only mode in the system, whereas at $\vartheta=20^\circ$ integer multiples of the initial mode are clearly visible. Due to the symmetry of the initial condition, odd multiples of the initial wave-vector are favored over the even multiples. The intensities of the odd modes are well fit by an exponential  $e^{-c\Delta t |\mathrm{k}|/|\mathrm{k}_{\mathrm{ini}}|}$ with $c\approx 0.37~\mathrm{s}^{-1}$  (dotted green line). The parameter $c$ combines both the scattering rate and damping of the spin waves which can not be trivially disentangled. \label{singleq_freq}}
\end{figure}

To get some insight into the wavenumber redistribution we consider the time evolution of a plane wave excitation:
\begin{align}
   \mathbf{m}(x,y) &= m_0~\hat{\mathbf{e}}_\mathrm{x} - \sin \vartheta \sin  2 \pi \mathbf{r} \cdot \mathbf{k}_\mathrm{ini}~\hat{\mathbf{e}}_\mathrm{y} \ , \label{initial_singleq}
\end{align}
with $\mathbf{k}_\mathrm{ini}=\left( 0.01, 0.005 \right) \ \mu\mathrm{m}^{-1}$. As before, $m_0$ is a normalization constant ensuring $|\mathbf{m}|=1$ and $\vartheta$ governs the amplitude of the sinusoidal modulation of the magnetization. In k-space, this mode gives two sharp peaks at $\pm \mathbf{k}_\mathrm{ini}$. In Fig.\ \ref{singleq_freq} we show the time-average ($\Delta t = 1.176$ ns) of the wavenumber distribution along $\mathbf{k}_\mathrm{ini}$ for two initial excitations following Eq. \ref{initial_singleq}, varying $\vartheta$.
We find that the distribution does not change in time for $\vartheta = 1^\circ$. However, when we increase $\vartheta$ to $20^\circ$, modes with integer multiples of $|\mathbf{k}_\mathrm{ini}|$ are present. Our method does not allow us to study the interaction that leads to the wavenumber multiplication in detail, but it can be generally understood as an inelastic scattering process.

\begin{figure}[hbt]
\centering
\subfloat[\label{ellip_Y_theta40}]{
   \includegraphics[width=.4\textwidth]{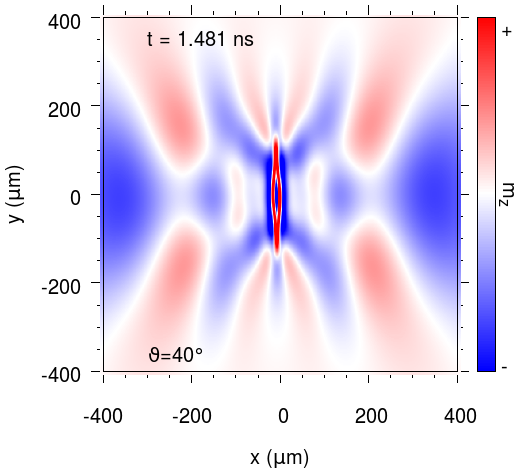} 
} \qquad
\subfloat[\label{ellip_X_theta40}]{
   \includegraphics[width=.4\textwidth]{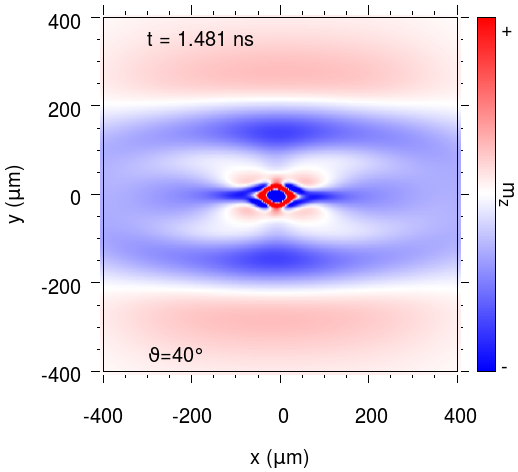}
}
\caption{Two snapshots at $t = 1.5$ ns of the spin wave pattern for elliptical initial excitations as in Fig. \ref{sw_compare_to_exp} and $\vartheta=40^\circ$ with: \protect\subref{ellip_Y_theta40}  $x_0= 35 \mu$m, $y_0=140 \mu$m; \protect\subref{ellip_X_theta40} $x_0 =140 \mu $m, $y_0=35 \mu$m. The color coding is scaled to one quarter of the initial amplitude of the tilt $\vartheta$ to aid comparison with Fig. \ref{sw_compare_to_exp} and highlight the spacial features. For these large values of $\vartheta$, the spin waves power accumulates in the center of the beam spot. \label{nonlin_ellip}}
\end{figure}

As mentioned in the introduction, previous studies of spin waves in the nonlinear regime have been performed. \cite{boyle1996nonlinear,PhysRevLett.81.3769,Demokritov2001441} The observed nonlinear effects were interpreted using the nonlinear Schr\"odinger equation (NSE) for spin waves. \citep{whitham2011linear,karpman2013non, stancil2009spin, zvezdin1983contribution} This equation can be derived when one expands the spin wave dispersion around a wave-vector $\mathbf{k}_0$ for small variations in the wavenumber and the spin wave amplitude. For a spectrally narrow wave packet, the (2 dimensional) NSE predicts (quasi) stable propagating solutions where the spin wave packet does not change its shape, a spin wave bullet.\cite{buttner2000linear}\\
\newpage
Our results of nonlinearities of BVMSW cannot be explained in the framework of the NSE. Firstly, the optical excitations that we study create spectrally broad excitations where a carrier wave-vector $\mathbf{k}_0$ cannot be identified. Secondly, the optical excitation modeled in this paper is centered at $k=0$ and due to the discontinuous derivative of the spin wave dispersion, an expansion of the dispersion around this point is ill-defined. 
The self-focusing spin waves we observe have a qualitatively different behavior than those described by the NSE, since they accumulate in the same location for all times. We conclude that the nonlinear phenomena observed are of different nature that those described in previous studies. \\
\newpage
For completeness, we have calculated the spin wave pattern for the two elliptical optical pulses used in Ref. \onlinecite{satoh2012directional} for large initial excitation. Fig. \ref{nonlin_ellip} shows two snapshots of the spin wave pattern at $t=1.5$ ns for $\vartheta=40^\circ$. The self-focusing effect is also clearly visible when compared to the linear ($\vartheta=1^\circ$) case shown in Fig. \ref{sw_compare_to_exp}.  The shape of the focal point is different for both elliptical excitations and is determined by the initial conditions and the spin wave dispersion.

\newpage

\section{Discussion \label{discuss} }
In this paper, we have presented numerical results on volume mode spin wave propagation in \ce{Gd4/3Yb2/3BiFe5O12}. We modeled the effect of a femtosecond pulse of circularly polarized light by an instantaneous tilt in the saturation magnetization over an angle $\vartheta$ in the direction perpendicular to the applied field and the equilibrium magnetization. Our simulations were performed in the uniform mode approximation and reproduce experimental results in the linear regime with very good agreement.
For large values of $\vartheta$, we predict that nonlinear effects can lead to the appearance of small wavelength modes which are not present in the wavenumber distribution of the pump pulse. In this case, the spin wave power at the location of the optical excitation remains high. The observed nonlinear effects are qualitatively different from those studied in previous works.

Our results for small $\vartheta$ coincide with experimental and computational results in the linear case \cite{satoh2012directional} and predict self-focusing behavior and the generation of small wavelength modes for larger $\vartheta$. Some comments on the experimental feasibility of these large $\vartheta$ are in order.

A simple calculation shows that one requires a magnetic field of $\sim 0.83$ T for 120 fs to tilt a single spin over an angle of $1^\circ$.\cite{stohr2006magnetism} Since the pulse duration is three orders of magnitude shorter than the timescale of the volume mode spin wave dynamics, longer pulses increase the tilt angle without affecting the validity of our approach. Furthermore, since the IFE does not rely on absorption of the incident light, a larger tilt angle could be achieved by increasing the fluence of the pump pulse. For instance in a DyFeO$_3$ garnet, a pulse of 200 fs with a fluence of $500$ mJ/cm$^2$ induced fields of $5$ T.\cite{kimel2005ultrafast} Alternatively, the magnetization could be manipulated via the magnetic component of the pump pulse. Ref.~\onlinecite{kampfrath2011coherent} reported a tilt of $0.4^\circ$ with a driving field of THz pulses peaking at $0.13$ T. It has been suggested in Ref.~\onlinecite{sell2008phase} that by using higher-frequency optical pulses $\sim 10$ T fields could be achieved, bringing the system in the nonlinear regime.
\newpage
In the nonlinear regime, the self-focusing leads to large amplitude precession of the magnetization at the center of the pulse spot. The shape of the spin wave focus is determined by the spin wave dispersion $\omega (\mathbf{k})$ and the shape of the incident beam. It is at least an order of magnitude smaller than the dimensions of the provided optical pulse. We hope that our predictions will stimulate experiments in the  nonlinear regime and the study of self-focusing of spin waves in the two dimensional plane. Control of the spin wave flow in combination with nonlinear behavior could be useful in technological applications where a strong precession affects a local magnetic structure, such as in switching devices. 

\begin{acknowledgments}
The authors want to acknowledge A.\ Kimel for his contributions at the start of the project.  We want to thank T.\ Satoh and K.\ Shen for useful discussions and our reviewers for valuable insight and criticism. This work is part of the research program of the Foundation for Fundamental Research on Matter (FOM), which is part of the Netherlands Organization for Scientific Research (NWO). Furthermore, the work is supported by European Research Council (ERC) Advanced Grants No.\ 338957 FEMTO/NANO and No.\ 339813EXCHANGE.
\end{acknowledgments}

\end{document}